\documentstyle[twoside,fleqn,espcrc2]{article}
\newcommand{\be}{\begin{equation}} 
\newcommand{\ee}{\end{equation}} \newcommand{\ba}{\begin{eqnarray}}
\newcommand{\ea}{\end{eqnarray}} \newcommand{\la}{\label}

\newcommand{\AmS}{{\protect\the\textfont2
  A\kern-.1667em\lower.5ex\hbox{M}\kern-.125emS}}

\title{Old Puzzles \hfill{\large CERN-TH/2000-254}}

\author{Christof Schmidhuber
        \thanks{Based on an invited talk at the Swiss Pysical Society meeting in Montreux, March 2000. Extended Feb. 2002.}}

\input epsf
\begin{document}

\begin{abstract}

\noindent

\end{abstract}

\maketitle

\section{Introduction}

We all know that  matter such as the apple in figure 1 is made of molecules,
molecules are made of atoms, and atoms are made of even smaller particles.
It is a very old question what the most fundamental building blocks of matter
are.

What we observe in the world is not only matter -- there are also other
things, shown in figure 1, such as
light, and forces such as the electromagnetic or the gravitational
force. All this
is embedded in a three--dimensional space, and there is also one time
dimension. It is another very old question where all this comes from.

 \begin{figure}[htb]
 \vspace{9pt}
\vskip5cm
 \epsffile[1 1 0 0]{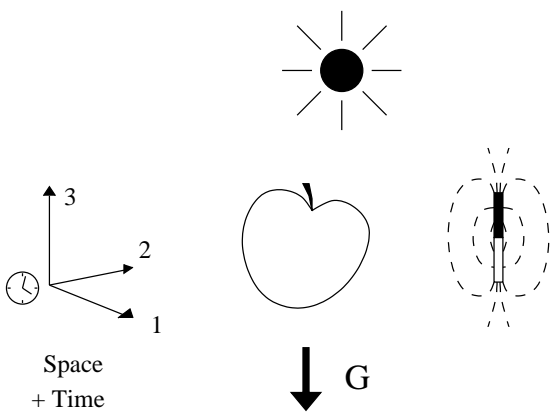}
\caption{What we observe}
\end{figure}

In my talk I first want to briefly review how far we have come in
answering those
old questions. I will begin with the things we know, which is 
the
``Standard Model'', and then talk about the things that 
we
can guess, which is
superstring theory. 
After this review I will emphasise a 
key point at which our understanding of superstring
theory presently stops: the problem of supersymmetry
breaking and the cosmological constant. I will also explain
in which direction I imagine a way out. 

In a separate note \cite{schm} I will discuss
my idea about what strings are made of.

\section{The Standard Model}

So what are the things we know?
We know that the atoms that the apple is made of consist of electrons
and nuclei,
the nuclei are made of protons and neutrons, and protons and neutrons
are made of
three quarks each. The quarks come in three different states called
``colors'' --
red, green and blue. Electrons and quarks are particles of spin
${1\over2}$.
There is another spin ${1\over2}$ particle that is harder to detect, the
neutrino.
Together with the electron
and two types of quarks, called $u$ and $d$, the neutrino forms a
`family' of quarks
and leptons (figure 2). Alltogether we know three such families.
So much for ``matter'', i.e., for spin ${1\over2}$ particles.

As for light, it is of course nothing but waves in the field of the
electromagnetic
force.
Quantum field theory describes forces parallel to matter by elementary
particles,
but with integer spin. In the case of electromagnetism we have a spin 1
particle,
the photon (figure 3).

Apart from the electromagnetic force there is also the ``weak force''.
The
corresponding
elementary particles of spin 1 are two W--bosons and one Z--bosons.
Among
other
things, they can turn the electron and the neutrino into each other.
More
precisely,
there is an $SU(2)$ symmetry associated with rotations in an internal complex
two--dimensional vector space. The electron and the neutrino transform in
the fundamental representation of this group and differ only by their
orientation in this vector space;
and the so-called "gauge bosons" W and Z
transform in the adjoint representation of
this symmetry group.

 \begin{figure}[htb]
 \vspace{9pt}
\vskip4cm
 \epsffile[1 1 0 0]{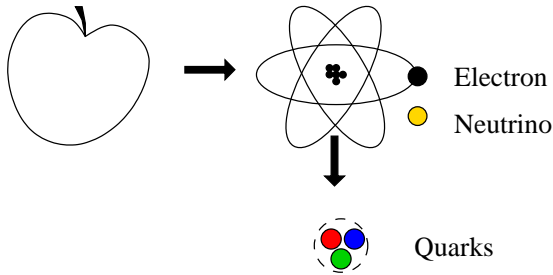}
\caption{Particles with spin 1/2}
\end{figure}

 \begin{figure}[htb]
 \vspace{9pt}
\vskip6.5cm
 \epsffile[1 1 0 0]{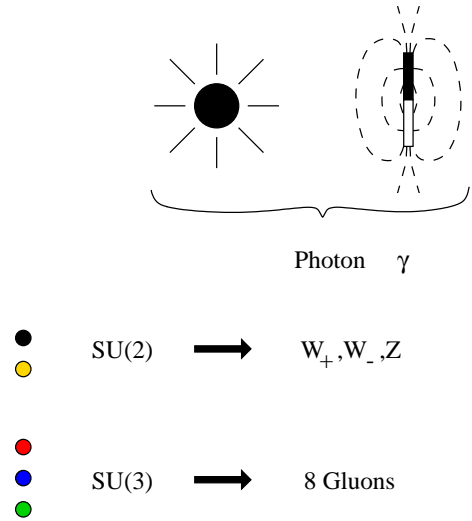}
\caption{Particles with spin 1}
\end{figure}

There is also an $SU(3)$ symmetry associated with rotations in the
3--dimensional complex
vector space generated by the colors of the quarks. The associated gauge bosons are 8 ``gluons''.
They mediate
the strong interactions (the nuclear force) and are described by Quantum
Chromodynamics (QCD). So alltogether, the known spin 1 particles are the
gauge bosons associated with a symmetry group
$$SU(3)\times SU(2)\times U(1)\ .$$
All the spin ${1\over2}$ and spin $1$ particles and their interactions
mentioned are described in very good agreement with experiment by the
Standard Model of Elementary Particle Physics,
which also contains a spin 0 particle -- the Higgs boson. The Higgs
boson is
held responsible for breaking the
$SU(3)\times SU(2)\times U(1)$ symmetry group
down to the observable $U(1)$ subgroup that corresponds to the
electromagnetic force.
It is the only particle that has not yet been observed directly in
experiments.

There are of course many interesting aspects and details of the Standard
Model which I will not go into since there are many other talks about
them at this conference. Instead,
let me move on to summarizing why we would like to go beyond the
Standard Model.

\section{Beyond the Standard Model}

Why are we not yet happy with the Standard Model?
First of all, there is of course the question why there should be so
many different
``elementary''  particles, rather than only a single one. Second, the
standard model
contains 18 free parameters -- the masses of these particles and the
strenghts of their
interactions -- that are not predicted by the theory but are adjusted by
hand to match the experimentally observed values. So clearly this is not
yet
a truly fundamental theory. And of course there are questions
such as ``why are there exactly three families of quarks and leptons?''
and ``why is the
gauge group exactly $SU(3)\times SU(2)\times U(1)$?''

Before continuing the list of open questions, let me mention that
there are ways
to adress the issues mentioned so far. For example, one can
assume that at very high energies, or equivalently at very small scales,
there is a ``Grand Unified Gauge Group'' such as $SO(10)$ or $E_6$,
that breaks down to the $SU(3)\times SU(2)\times U(1)$ subgroup
that is observable at lower energies. This would unifiy the different
gauge bosons
in the sense that they would just form one representation of the
grand unified group.
Likewise, all the different spin ${1\over2}$ particles in one family
would just be different members of a single representation, or multiplet,
 of the large
group.

One can also unify spin 1 particles and spin ${1\over2}$ particles by
introducing
a symmetry that interchanges them, ``supersymmetry''. With enough
supersymmetry,
the spin 0 Higgs boson can also be incorporated in such a ``supermultiplet''.

One can even estimate the scales, at which such unifications might
occur.
The three gauge groups $U(1)$, $SU(2)$ and $SU(3)$ come with three
coupling constants, $\alpha_1, \alpha_2$ and $\alpha_3$. In field
theory, such
coupling constants are not really constant, but change with the scale,
at which
scattering experiments are performed. By extrapolating from their
know values at large distance scales (low energy scales)  down to small
distance scales
(high energy scales), one can see at what scales they meet.

In figure 4, the scales run from the size of the atom ($10^{-10} m$)
through the size of the nucleus ($10^{-15} m$)  and the scale of
electroweak symmetry breaking (roughly $10^{-18} m$), which is about as far as
we can
see with accelerators, down to the
the Planck scale (a fraction of $10^{-34} m$), where the strength of gravity
becomes
comparable with the strength of the other forces.
Assuming that nothing else happens in between, the three coupling
constants
appear to unify at a scale (the ``GUT scale'') that is not much larger than
the Planck scale.

Actually, the three coupling constants do not meet exactly in the
non--supersymmetric
standard model. But if we assume that supersymmetry is restored at
distance scales
not much below the scale of electroweak symmetry breaking, then the
couplings
{\it do} meet within their error bars, at a scale only a few hundred times
as large as the Planck scale. What we do not yet understand is the large
``hierarchy'' of 16 orders of magnitude between the
Planck scale
and the scale of electroweak symmetry breaking.

 \begin{figure}[htb]
 \vspace{9pt}
\vskip9.8cm
 \epsffile[1 1 0 0]{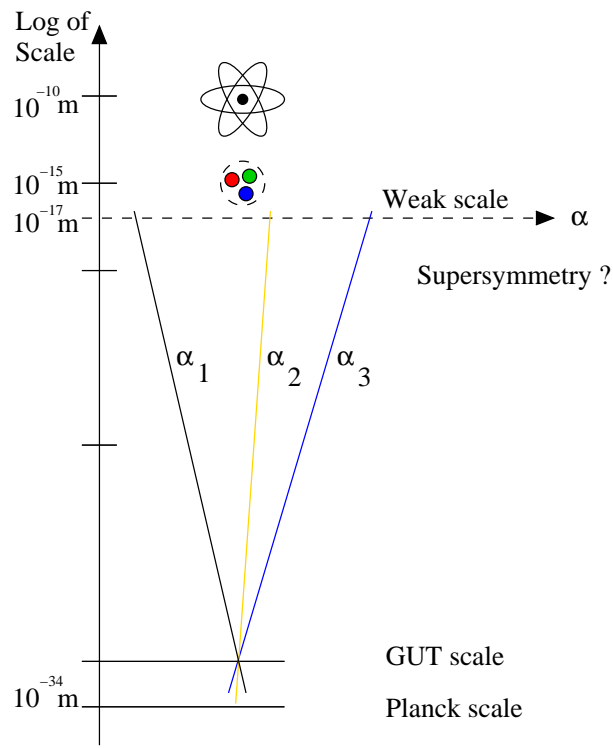}
\caption{Scales and running coupling constants}
\end{figure}

While grand unification and supersymmetry may be promising first steps,
they are not sufficient to adress two more fundamental
shortcomings
of the standard model. First of all, gravity doesn't seem to fit into
the framework.
The gauge boson of gravity would be a spin 2 particle, the graviton, and
there just isn't a consistent
quantum field theory of spin 2
particles, in the
sense of a renormalizable, unitary continuum theory. In the context of gravity,
another complete mystery that must be explained by a fundamental theory
is why the cosmological constant is so small after supersymmetry breaking.

Second, questions such as ``why is space--time four--dimensional?''
cannot
even be asked within the framework of the Standard
Model, which is a priori formulated in four space--time dimensions.

\section{Superstrings}

If there is no field theory of pointlike gravitons, the next thing
that one
could try after points are lines (called strings), surfaces (called membranes),
or higher-dimensional
extended objects.

Let me first say why I will only talk about strings here.
The reason is - and this has been understood only recently -
that whatever theory of extended objects we start with, inasfar
as the theory is consistent it is always just another
formulation of string theory.
E.g., there is the beginning of a theory of
membranes, the 11--dimensional supermembrane theory. But it has
been recognized that this membrane theory is in fact
equivalent to string theory in the sense of strong--weak coupling
duality.
This is basically a much more elaborated version of the more familiar
dualities of the two--dimensional Ising model (Kramers-Wannier duality)
or of supersymmetric Yang-Mills theory (Olive-Montonen duality):
the elementary excitations of the strong coupling limit (supermembranes) are
solitons of the weak coupling limit (superstring theory), and vice versa.
                                                              
So let us focus on string theory \cite{superstrings}. 
In string theory, all the different elementary
particles including the graviton are assumed
to be just different  excitation modes of a single fundamental
object, the superstring.
It can indeed be shown that this leads to a consistent quantum theory
that
includes, in particular, the concepts of grand unification and supersymmetry.
Superstring theory has only a single free parameter, the string tension
${1\over\alpha'}$.
And not even the space--time dimension is fixed in the theory, but
is a property of its ground state.

Then why are we still not happy?
Apart from the obvious question what strings are made of (we will come back to
it later),
the big problem with string theory is of course that it is so hard to
test whether
the theory is right or wrong. First of all, it seems unlikely that we
will ever
see the strings directly in experiments. It seems natural to assume that
their size, the ``string scale'' (called ${\sqrt{\alpha'}}$),  
is of the order of
the
Planck scale, and accelerators powerful enough to resolve such tiny
scales
would have to have galactic proportions.

One could still start from the hypothesis that the theory is correct,
and try to
``calculate it
down'' to low energies: what are its
predictions for the number of families of quarks and leptons, for the
gauge group,
for particle masses, and for the space--time dimension? Do they agree
with the
Standard Model?

Let me indicate briefly how this is done by discussing scattering
amplitudes. To simplify things I will pretend 
that spacetime has Euclidean signature. The scattering amplitudes can
be translated to Minkowskian signature via a
Wick rotation.   

Computing scattering amplitudes in Euclidean string theory is very much
like computing the energy of a soap film that is spanned, e.g., between
the three boundaries in figure 5. The soap film represents the
world--sheet that is swept out by the string as it
moves through its embedding space, and the boundaries represent, in this case, one
incoming
and two outgoing strings. One first finds the minimal area
surface that ends on the boundaries and computes its area $A$.
The energy of the soap film is $A$ times the tension ${1\over\alpha'}$.
Very roughly, with Minkowskian signature and
in the limit where the three strings at the boundaries
become very small,
the minimal area surface looks like an ordinary Feynman diagram of the
Standard Model, such as that for the decay of a $W^-$ boson into an
electron and an
antineutrino. I am simplifying a little bit, but basically
our soap film calculation  should in this limit yield a string theory
prediction
for the decay rate and for the associated weak coupling constant.

 \begin{figure}[htb]
 \vspace{9pt}
\vskip3.5cm
 \epsffile[1 1 0 0]{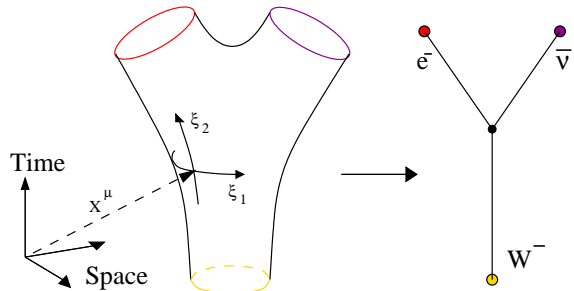}
\caption{String world-sheets and soap films}
\end{figure}

As I said I am simplifying things. The soap film calculation is what
one would do
for classical
strings, where the string world--sheet is a sharp
surface --
just like the world--line of a classical particle is a sharp line. In
quantum mechanics,
the string world--sheet fluctuates, though. To deal with this, one
labels the world--sheet
by coordinates $\xi_1,\xi_2$. One then parametrizes the world--sheet by
embedding coordinates $x^\mu(\xi)$, so that the space--time coordinates
$x^\mu$
become two--dimensional fields that live on the world sheet. Quantum
fluctuations
of the world--sheet are taken into account by
introducing a path integral  over the coordinates and weighing it with
$e$ to the
minus the ``soap film energy''. This yields $e$ to the minus the free
energy $F$:
$$\exp\{-F\}\ =\ \int {\cal D}x^\mu(\xi)\ \exp\{-{1\over\alpha'}A(x)\}\ .$$
Trying to do this path integral $\int {\cal D}x^\mu$
 leads one into the fascinating area of
two--dimensional
conformal field theory. The two-dimensional fields are the coordinates
$x^\mu(\xi)$, and $\alpha'$ plays the role of $\hbar$.

I have no time to
review two-dimensional field theory here, but I should take the time to
point out right from the beginning its crucial limitation: two dimensional
field theory can describe only {\it perturbative} string theory.

The diagrams in figure 5 are tree diagrams; they are computed from
 two-dimensional
field theory on the shown genus zero surface
with holes (the incoming or outgoing
strings). g-Loop diagrams correspond to surfaces of genus $g$. They are
suppressed by the $g$th power of the string coupling constant $\kappa^{2}$.
Thus the loop expansion of  particle theory becomes the genus expansion
of two-dimensional field theory (figure 6).

 \begin{figure}[htb]
 \vspace{9pt}
\vskip4cm
 \epsffile[1 1 0 0]{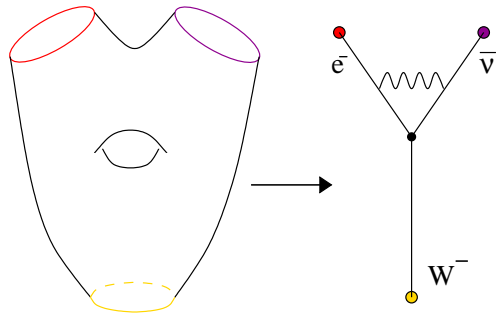}
\caption{1-loop scattering amplitudes from two-dimensional field theory on a torus}
\end{figure}

 But what about nonperturbative effects that are invisible
in the perturbation expansion in $\kappa$?
Those cannot be seen in two-dimensional field theory on a surface of any genus. That this is a crucial
limitation is clear from a comparison with QCD: just like
important phenomena in QCD such as confinement and chiral symmetry breaking
are invisible in its loop expansion, we must expect to miss
crucial aspects of string theory in two-dimensional field theory. We will
come back to this limitaion.
                                                                   
To conclude, there is a technical problem in calculating the
string theory predictions for the ``real world'', i.e. for the observable
particle physics at low energy scales (up to TeV): our present technology
allows us to only compute these predictions perturbatively in the
string coupling constant $\kappa$. Although recently discovered strong-weak
coupling dualities allow us to extend the range of validity from weak
coupling to various
strong coupling limits, the most interesting range of intermediate coupling
is presently out of reach.

\section{Perturbative string theory predictions}

After these words of caution, let me move on to
the results anyway -- the perturbative string theory predictions for the
observable world.

The good news is that gravity comes out right. One really recovers Einstein's theory of
general relativity
as part of the low--energy limit, so this is a big success.

The bad news is that 
supersymmetry is unbroken in perturbative string theory.
So Einstein's gravity is really part of supergravity. This disagrees with the
real world - we do not observe massless superpartners of the graviton.
Moreover, it brings with it a long list of
other problems. As is typical for supersymmetric theories, the ground
state is degenerate. There are literally millions of distinct perturbative
ground states with exactly the same energy (namely zero), and each of
them makes a different prediction for the low-energy gauge group, the number
of families, the number of scalar (Higgs) fields, etc.

Certainly the simplest ones of these ground states do not resemble the
real world. 
Their gauge group comes out to be not $SU(3)\times SU(2)\times U(1)$,
but either $SO(32)$ or $E_8\times E_8$. The number of families of quarks
and leptons comes out to be  not three but zero. But the most
embarassing thing that comes
out wrong is the number of space--time dimensions: it comes out to be 10! 

Many - but not all - of the different ground states can be geometrically
interpreted as compactifications of the simple 10-dimensional ones.
When you roll up a sheet of paper, it looks
one--dimensional rather than two--dimensional from the distance (figure 7).
Likewise, six of the ten dimensions of string theory might be curled up
into
a tiny compact manifold (it  turns out that it must be a Calabi--Yau
manifold)
such that the world looks 4--dimensional at scales much larger than the
string
scale. It can be seen that in this way one can also break the gauge
group down
to $SU(3)\times SU(2)\times U(1)$, and one can also obtain chiral families of
quarks
and leptons.

 \begin{figure}[htb]
 \vspace{9pt}
\vskip2cm
 \epsffile[1 1 0 0]{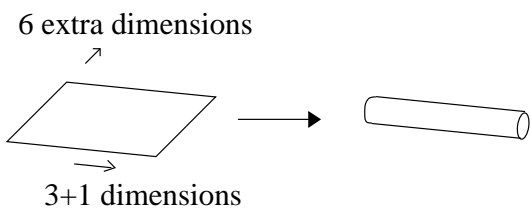}
\caption{4 dimensions from 10 dimensions}
\end{figure}

Millions of different ground states result because there are millions of
topologically
distinct ways to compactify six dimensions.
In figure 7, just a few of these
compactifications
are plotted \cite{klemm}. Each point represents
one possible compactification. What is plotted is the number of Higgs fields
(up to 480) that they
predict
against the number of families of chiral fermions
 that they predict (also up to 480!). 
Do we live on one of
these
points? Well, even if we did find a point that agrees with the observable 
world in terms of gauge group, number of Higgs fields and number of lepton
families,
we would not be convinced that string theory is correct. By picking a
different point,
we could ``predict'' pretty much any low energy physics we like -- or
in other words, we really cannot predict anything.

 \begin{figure}[htb]
 \vspace{9pt}
\vskip6cm
 \epsffile[1 1 0 0]{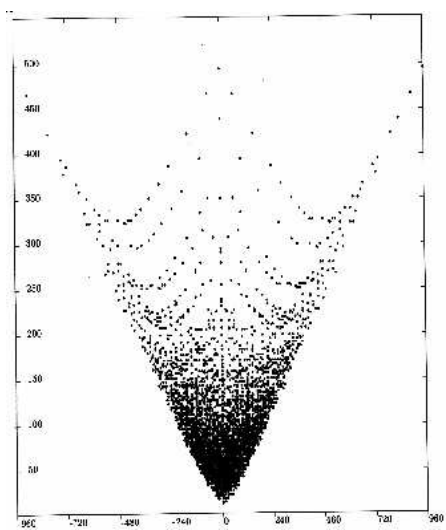}
\caption{Some of the many string vacua}
\end{figure}

The reason for this lack of predictive power of perturbative string theory
has already been mentioned: supersymmetry is unbroken perturbatively. Supersymmetry breaking
must be a {\it nonperturbative} effect in string theory that is missed by
the genus expansion of string amplitudes. Once
supersymmetry is broken, the huge degeneracy of the string ground state
can be expected to be lifted. So there should be a true vacuum of string
theory after all, and it should lead to  unique predictions.

The big obstacle to making contact between superstring theory
and the real world is then to understand how supersymmetry is
broken non-perturbatively.

\section{The cosmological constant}
  
The big {\it clue} to understanding
supersymmetry breaking must be the cosmological constant: mysteriously,
nature seems to manage to break supersymmetry without generating
a huge cosmological constant. We first make a brief digression
to recall what is so mysterious about this cosmological constant problem.
                                              
The energy density $\rho$ of the vacuum enters Einstein's
equations in the form of an effective cosmological constant
$\lambda$:
\ba R_{\mu\nu}- {1\over2} g_{\mu\nu} R = \lambda g_{\mu\nu}
\ \ \hbox{with}\ \ \lambda\ =
8\pi G \rho\ ,\la{anna}\ea
where $G$ is Newton's constant. Such a cosmological constant $\lambda$
curves four--dimensional space--time, giving it a curvature
radius $R_{curv}$ of order $\lambda^{-{1\over2}}$.
Only part
 of the curvature of the universe is due to the cosmological constant;
other contributions come from visible and dark matter.
This is accounted for by a factor
$3\Omega_\Lambda$:
$$3\Omega_\Lambda\ R_{curv}^{-2}\ \sim\ \lambda\ .$$
$\Omega_\Lambda$ seems to have been
measured to be $\Omega_\Lambda\sim{2\over3}$ \cite{bah}.
For a spacially flat universe, which is the case that
seems to be realized in nature, this curvature radius implies an
expanding universe with metric
\ba ds^2\sim - dt^2 + e^{Ht} d{\vec x}^2\ \ \hbox{with}\ \
3\Omega_\Lambda H^2\sim \lambda .\la{hugo}\ea
The Hubble constant $H$ in our universe is of order
$$H\ \sim\ 10^{-33}eV\ \sim \ (10^{26}m)^{-1}\ .$$
The cosmological constant problem is the question why this
is so small, given that we expect  much larger contributions
to the energy density of the vacuum from Standard Model physics.
Let us briefly summarize why we expect such contributions.

First of all, there are classical effects.
E.g., in the course of electroweak symmetry breaking the Higgs field
is supposed to roll down its potential, thereby changing the
vacuum energy density by an amount of order $(TeV)^4$.
This would give a curvature radius of the universe in the millimeter
range, which is not what we observe. Of course, the Higgs potential
could be shifted such that its Minimum is at zero. Equivalently,
 a bare cosmological
constant could be added in (\ref{anna}) that exactly cancels
the cosmological constant induced by $\rho$. But this is just the fine--tuning
problem: why should the bare cosmological constant be fine--tuned to
exactly cancel the contributions from later phase transitions?

Even if we do find a reason why the minimum of the Higgs potential
should be zero, the problem does not go away. There are other condensates
in the Standard Model, such as the chiral and gluon condensates in
QCD after chiral symmetry breaking. They should contribute an amount
of order $(\Lambda_{QCD})^4$ to the cosmological constant, which would
translate into a curvature radius of the universe in the kilometer range.

And even if one did not believe in chiral and gluon condensates,
there would still remain what is perhaps the most mysterious aspect
of the problem: the cosmological constant does not seem to
receive contributions from the zero--point energy of the Standard Model
fields. A harmonic  oscillator has a ground state energy ${1\over2}\hbar
\omega$. In a field theory we have one oscillator for each
momentum $k$ with frequency
$$\omega\ =\ \sqrt{k^2+m^2}\ .$$
In the case of the photon (with $m=0$), integrating
over $k$ gives, for each of the two polarizations, a divergent  contribution
\ba\lambda\ \ =\ 8\pi G\ \int_{|k|\le\Lambda}{d^3k\over(2\pi)^3}{1\over2}
\hbar\omega\ \sim\ {1\over2\pi} {\Lambda^4\over m_{Pl}^2}\la{beate}\ea
to the cosmological constant, where $\hbar G\ =\ m_{Pl}^{-2}$ and
$\Lambda$ is a large--momentum cutoff. How large does
this cutoff $\Lambda$ have to be?
(\ref{beate},\ref{hugo}) tell us that $\Lambda$ is, on a logarithmic scale, half--way
between the Planck mass and the Hubble constant. 

\ba {H\over\Lambda}\ \ \ \sim\ \ \ {\Lambda\over m_{Pl}}.\la{haribo}\ea

It is straightforward to calculate that $\Lambda$ is several
milli-eV, corresponding to a Compton wave length in the micrometer range.
This is even larger than the wave--length of visible light: we can
see with our bare eyes that there is no such cutoff in nature!            

What {\it can} explain a zero cosmological constant is
supersymmetry. In supersymmetric theories, the contributions from
bosons and fermions to $\lambda$ have opposite signs and
cancel. But in order to explain the smallness of the Hubble constant,
supersymmetry would have to remain unbroken at least up to
scales of the order of the above cutoff $\Lambda$, i.e. up to milli-eV
scales (see figure 1). 

In other words, the cosmological 
constant that seems to have been observed
 could be produced by a supermultiplet of
particles with mass splittings of order milli-$eV$
(at least if the supertrace of the mass matrix of the supermultiplet vanishes).
But unbroken supersymmetry up to milli-eV scales is of course ruled out 
for the Standard model: 
there is no superpartner
of the photon with mass in the milli-eV range. If there was, the
energy levels in atoms would be very different (if atoms were stable at all).
So this is the dilemma.

\section{Do we live on a soliton?}

We should probably 
be grateful for this dilemma.
The miracle of the nearly vanishing cosmological constant
should be a crucial hint as to {\it how} nature breaks supersymmetry.
Let me now tell you what I personally think that the smallness of the cosmological constant hints at:
that we live on a non-supersymmetric soliton of superstring theory.

You might be familiar with extended solitons in
solid state physics such as vortex lines in superconductors.
10-dimensional string theory also contains a variety of extended solitonic
objects,
so--called $p$--branes. A $p$--brane has $p$ space-- and one time
dimension.
There are i.p.  3--branes (figure 8). Each point on it is surrounded by
an $(n-1)$--sphere,
if there are $n$ dimensions transverse to the brane (in our case, $n=6$,
but it is
useful to consider the more general case). It can be shown that there
are
gauge fields $A^\mu$, chiral fermions $\psi$ and scalar fields $\phi$ that live only on the
3--brane.
These fields might make up the particles of the standard model,
if we identify the 3--brane with the observable universe. Gravity,
however, is not restricted to the brane in these models.
Instead it lives all over the
$(4+n)$--dimensional bulk.

 \begin{figure}[htb]
 \vspace{9pt}
\vskip5cm
 \epsffile[1 1 0 0]{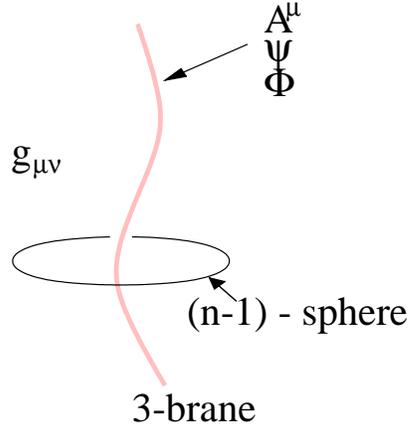}
\caption{3-branes embedded in $4+n$ dimensions.}
\end{figure}

Imagine now that we live on a stable {\it non-supersymmetric} soliton
- a 3+1 dimensional brane inside the 10-dimensional space
of superstring theory;
so supersymmetry is broken only because of this
brane. 
Why might this help solve the cosmological constant problem?

I will sketch the ideas of \cite{schmi}.
As we noted before, at
least if the recent measurements of the cosmological constant
can be trusted, the size of the cosmological constant is such
that it could be produced by a supermultiplet of particles with
supersymmetry being broken in the milli--$eV$ range.
Whatever these particles are, they must be fundamentally different
from the Standard Model particles, because they contribute to the
cosmological constant, while the Standard Model particles obviously
don't.

The non--supersymmetric brane world scenario provides precisely such a fundamental
difference between the Standard Model particles, which live
on the brane, and other particles (the supergravity particles),
which live in the bulk. So the idea is that the cosmological constant
is produced only by the supergravity multiplet. 

But why should the Standard Model particles not contribute
to the cosmological constant at all? Well, the scenario does in fact
provide a mechanism
 which could at least in
principle do the job of soaking up the Standard Model vacuum energy:
if the Standard Model is confined to a brane, then it 
is conceivable that the resulting brane cosmological constant
only curves the space transverse to the brane,
but not the space parallel to the brane \cite{rub} --
so the curvature radius of the four-dimensional universe
could still be huge.

It is very nontrivial and striking that
the supersymmetry breaking scale $\Lambda$ of order milli--$eV$ in the gravity sector
comes out naturally, if one assumes Standard Model supersymmetry
breaking in the $TeV$ range (as required for the running coupling constants to meet, as mentioned earlier). One then expects e.g. a gravitino mass of order
$$m_{3/2}\ \sim\ {{(TeV)^2\over m_{Pl}}}\ \sim\ \Lambda\ \sim\ \hbox{milli-}eV\ .$$
Together with (\ref{haribo}),
one in fact predicts a 
relation between four vastly different scales in physics \cite{schmi}:
the Planck scale, the scale of supersymmetry breaking in the
Standard Model, the inverse gravitino mass and the inverse Hubble expansion
rate of the universe, whose square is of the
same order $(10^{-33} eV)^2$ as the value of the cosmological constant
that seems to have been recently observed \cite{bah}.
This predicted relation is plotted in figure 9, which extends
figure 4 up to cosmic scales. 
The solid horizontal lines are predicted to be equally spaced on a logarithmic
scale, with step size roughly 15.5.

 \begin{figure}[htb]
 \vspace{9pt}
\vskip8cm
 \epsffile[1 1 0 0]{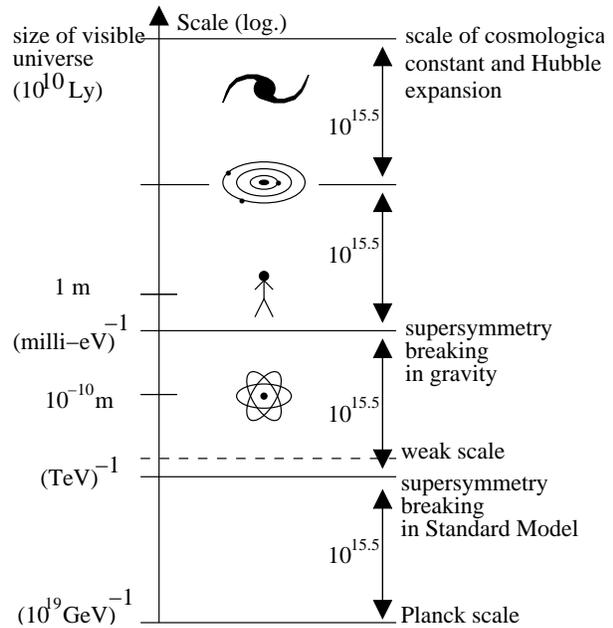}
\caption{Predicted ratio of scales (to convert
electron volt into meters: 
$(1 eV)^{-1}\sim2\cdot 10^{-7}m$.)}
\end{figure}

While there are certainly unresolved issues in this scenario (see, e.g.,
second reference in \cite{schmi}), 
at least to the author it feels
like pieces of a puzzle beginning to fall into place.
But perhaps the best feature of this scenario is that it can be tested 
experimentally: it predicts gravitinos and other superpartners of the graviton with masses
in the $milli-eV$ range.
The precise masses of the superpartners of the graviton
 depend on the detailed non-BPS soliton solution, so 
the spectrum
of supergravity masses should in principle be a window through which we
can probe what kinds of branes our string compactification contains.

Masses in the $milli-eV$ range
 translate into a hypothetical
 gravitino or dilaton with Compton wavelength in the
micrometer range. 
This is just at the border of not yet
being ruled out by experiment,
and it could be checked by short-distance 
measurements of gravity 
in the not-so-far future. 

\vskip1cm

\section{Conclusion}

To summarize, the observation of apples, light and electromagnetic
forces leads us to the Standard Model. The observation of gravity leads
us to superstrings. 

To compute the low-energy predictions
of superstring theory and compare them with observation,
a crucial step is to
understand how supersymmetry is broken in string theory.

The crucial hint should be the smallness of the cosmological constant:
its observed size is precisely such that it could be produced by the vacuum energy
of the supergravity sector, if the
observable universe was a four-dimensional
non-supersymmetric soliton inside the ten-dimensional space-time 
of superstring theory.

This scenario
predicts gravitinos and other superpartners of the graviton
with masses in the milli-$eV$ range. They could be observed 
experimentally through short-distance measurements of gravity in the micrometer range.

\end{document}